\documentclass[10pt, prl,aps,twocolumn,%
showpacs,preprintnumbers,amsmath,amssymb]{revtex4-2}
\usepackage[utf8]{inputenc}
\usepackage{graphicx}
\usepackage{bm}
\usepackage{physics}
\usepackage{siunitx}
\usepackage[breaklinks, colorlinks=true, citecolor=blue]{hyperref}

\newcommand{\br}{\bm{r}}
\newcommand{\bS}{\bm{S}}

\newcommand{\bK}{\bm{K}}
\newcommand{\bp}{\bm{p}}
\newcommand{\bDelta}{\bm{\Delta}}
\newcommand{\bB}{\bm{B}}
\newcommand{\bBo}{\bm{B_\omega}}
\newcommand{\Q}{q_\text{top}}
\newcommand{\FQ}{F_Q}
\newcommand{\topsus}{\chi_\text{top}}
\newcommand{\Lag}{\mathcal{L}}

\begin{document}

\title{Polarized Nucleon as a Topological Dipole}

\author{Kenji Fukushima}
\email{fuku@nt.phys.s.u-tokyo.ac.jp}
\affiliation{Department of Physics, The University of Tokyo, 
  7-3-1 Hongo, Bunkyo-ku, Tokyo 113-0033, Japan}

\author{Tomoya Uji}
\email{uji@nt.phys.s.u-tokyo.ac.jp}
\affiliation{Department of Physics, The University of Tokyo, 
  7-3-1 Hongo, Bunkyo-ku, Tokyo 113-0033, Japan}

\begin{abstract}
  We show that a polarized nucleon generically carries a dipole distribution of topological charge density.  This topological dipole follows robustly from the definition of the topological form factor and the pseudoscalar nature of the topological charge density.  The strength of the topological dipole is fixed, in the chiral limit, by the flavor-singlet axial charge.  We demonstrate the mechanism in a two-flavor chiral soliton model with vector mesons and the $U(1)_A$ anomaly, where the rotation of the soliton induces a singlet pseudoscalar profile and realizes the predicted dipole pattern.  We also discuss possible experimental probes through exclusive $\eta$, $\eta^\prime$ production and directed-flow-like pseudoscalar-meson asymmetries correlated with magnetic fields or vorticity in relativistic heavy-ion collisions.
\end{abstract}

\maketitle

\paragraph{Introduction:}
One of the most characteristic nonperturbative aspects of quantum chromodynamics (QCD) is the topology of gluon fields.~\cite{Belavin:1975fg, tHooft:1976snw, Schafer:1996wv}.  At the local level, this topology is characterized by the topological charge density, $\Q = \alpha_s/(4\pi)\Tr F_{\mu\nu}\widetilde F^{\mu\nu}$.  With the usual boundary conditions, the spacetime integral of $\Q$ gives the topological charge, or winding number, of the gauge-field configuration.  The same operator plays a central role in the $U(1)_A$ anomaly~\cite{Adler:1969gk, Bell:1969ts}.  It appears in the divergence of the flavor-singlet axial current,
\begin{equation}
    \partial_\mu j_5^\mu = 2N_f\Q + 2\sum_f m_f\bar \psi_f i\gamma_5 \psi_f.
    \label{eq:anomaly_intro}
\end{equation}
This identity ties gluon topology to two central questions in QCD.  In the pseudoscalar sector, it provides the standard resolution of the $U(1)_A$ problem by generating the anomalously large $\eta^\prime$ mass~\cite{Witten:1979vv, Veneziano:1979ec}.  In nucleon matrix elements, the same identity enters the interpretation of the flavor-singlet axial charge, $\Delta\Sigma$, measured in polarized deep-inelastic scattering (DIS)~\cite{Jaffe:1989jz}.

The connection between the chiral anomaly and the proton spin problem in Ref.~\cite{Jaffe:1989jz} has recently attracted renewed attention in high-energy spin physics, particularly in view of the Electron Ion Collider.  In polarized DIS, the anomaly has been argued to affect $g_1$ beyond its first moment, including in the Bjorken and Regge limits~\cite{Tarasov:2020cwl, Tarasov:2021yll, Tarasov:2025mvn}.  Related anomaly-induced structures have also been discussed in deeply virtual Compton scattering and generalized parton distributions (GPDs), where anomaly poles and their cancellation are encoded in the axial GPD sector~\cite{Bhattacharya:2022xxw, Bhattacharya:2023wvy, Bhattacharya:2024geo, Castelli:2024eza}.  These developments point to a broader program of using high-energy inclusive and exclusive processes to probe how QCD topology enters the spin structure of the nucleon.

These discussions are usually formulated in terms of light-cone correlation functions and their moments.  In this Letter, we ask a complementary question: what is the Breit-frame distribution encoded in the nucleon matrix element of the topological charge density?  Because $\Q$ is a pseudoscalar, a spin-$1/2$ target cannot support a spherically symmetric distribution of $\Q$.  With $\br$ measured from the nucleon center in the Breit frame and
$\hat{\br} = \br / r$, the allowed angular dependence is therefore
$\bS\cdot\hat{\br}$, corresponding to a topological dipole aligned
with the nucleon spin direction $\bS$.  We show that this dipole structure follows model-independently from the topological form factor, and that its dipole moment is related in the chiral limit to the singlet axial charge, $d_Q = - \Delta\Sigma / N_f$.  We then illustrate this structure in a chiral soliton model with the $U(1)_A$ anomaly, where the induced singlet pseudoscalar profile explicitly realizes the spin-induced topological dipole.

This dipole nature also suggests a new connection to the search for local $P$-odd topological structure in relativistic heavy-ion collisions.  In the conventional scenario, sphaleron transitions generate topological charge with a random sign event by event~\cite{Kharzeev:2007jp}.  The resulting chirality imbalance induces an electric current along the magnetic field, leading to charge separation with a random sign, which is known as the chiral magnetic effect (CME)~\cite{Fukushima:2008xe}.  Experimental searches have therefore focused on $P$-even correlators of charge separation, most notably the $\gamma$-correlator~\cite{Voloshin:2004vk}.  Extensive measurements at the Relativistic Heavy Ion Collider (RHIC) and the Large Hadron Collider, including isobar collisions, have made this a mature program, but the interpretation remains limited by large flow-induced and nonflow backgrounds~\cite{Kharzeev:2015znc, Zhao:2019hta, STAR:2021mii}.

By contrast, the spin-induced topological dipole proposed here provides a different way for local topological charge density to survive event averaging.  The spatial integral of the dipole vanishes, so no net topological charge is generated.  Nevertheless, if baryonic spin polarization is correlated with the magnetic field, $\bB$, the dipole orientations can be biased, producing an event-averaged local topological polarization, $\langle \Q(\br)\rangle_{\bB}\propto \hat{\bB}\cdot\hat{\br}$.  Here, $\langle\cdot\rangle_{\bB}$ denotes an event average after aligning the direction of $\bB$.  This would represent a nonzero local one-point function of $\Q$, rather than only an event-by-event fluctuation, and could be searched for through directed-flow-like azimuthal asymmetries in pseudoscalar meson production.
\vspace{0.5em}

\paragraph{Topological form factor and dipole density:}
The nucleon matrix element of $\Q$ is parametrized as~\cite{Jaffe:1989jz, Zahed:2022wae}
\begin{equation}
    \frac{1}{M_N}\mel{p^\prime, s^\prime}{\Q}{p, s} = \bar u^{(s')}(p^\prime) i\gamma_5 u^{(s)}(p)F_Q(t)
    \label{eq:FQ_def}
\end{equation}
with $\Delta^\mu = p^{\prime\mu} - p^\mu$ and $t = \Delta^2$.  The nucleon spinors are normalized as $\bar u^{(s)}(p)u^{(s)}(p) = 2M_N$.  We refer to $\FQ(t)$ as the topological form factor of the nucleon.

This form factor is related to the singlet axial form factors.  The nucleon matrix element of the flavor-singlet axial current, $j_5^\mu(x)=\sum_f \bar\psi_f(x)\gamma^\mu\gamma_5\psi_f(x)$, is parametrized as
\begin{equation}
\begin{split}
    &\mel{p^\prime, s^\prime}{j_5^\mu}{p, s} \\
    &\quad = \bar u^{s')}(p^\prime)\left[\gamma^\mu G_A^{(0)}(t) + \frac{\Delta^\mu}{2M_N}G_P^{(0)}(t)\right]\gamma_5 u^{(s)}(p)\,.
    \label{eq:singlet_axial_FF}
\end{split}
\end{equation}
By taking the divergence and using the anomalous Ward identity in the chiral limit, i.e., $\partial_\mu j_5^\mu = 2N_f\Q$, the topological form factor is related to the singlet axial form factors as~\cite{Jaffe:1989jz}
\begin{equation}
    2N_f\FQ(t) = 2G_A^{(0)}(t) + \frac{t}{2M_N^2}G_P^{(0)}(t)\,.
    \label{eq:FQ_GA_GP}
\end{equation}

Form factors are commonly used to construct the spatial structure of hadrons.  For example, the electromagnetic form factors give charge and magnetization distributions~\cite{Hofstadter:1958psf, Ernst:1960zza, Sachs:1962zzc}, the flavor-nonsinglet axial form factor
characterizes the spin distribution associated with the axial-vector current~\cite{Bernard:2001rs, Chen:2024oxx}, and the energy-momentum tensor form factors encode the mechanical distributions~\cite{Polyakov:2002yz, Polyakov:2018zvc}.  In the same spirit, the topological form factor defined in Eq.~\eqref{eq:FQ_def} provides access to the spatial distribution of the local topological charge density.

To convert $F_Q(t)$ to the spatial distribution, the Breit frame with $\Delta^\mu = (0, \bDelta)$ is the most convenient.  In the Breit-frame, the pseudoscalar spinor bilinear in Eq.~\eqref{eq:FQ_def} reduces as $\bar u^{(s')}(p^\prime)i\gamma_5 u^{(s)}(p) = - i\chi_{s^\prime}^\dagger\bm\sigma\cdot\bDelta\chi_s$, where $\chi_s$ is a two-component spinor.  Taking the spin-diagonal matrix element, $s^\prime = s$, and introducing the spin-polarization vector, $\bS = \chi_s^\dagger\bm\sigma\chi_s$, we express the Fourier transformation of Eq.~\eqref{eq:FQ_def} as
\begin{equation}
    \Q(\bS, \br) = \int\frac{d^3\bDelta}{(2\pi)^3} \,e^{-i\bDelta\cdot\br}\left[- i\bDelta\cdot\bS F_Q(- \bm{\Delta}^2)\right]\,.
    \label{eq:Q_FT}
\end{equation}
We note that the Breit-frame distribution corresponds to the probability density only in the non-relativistic limit, and relativistic light hadrons should involve subtle effects of localization and recoil~\cite{Lorce:2020onh, Jaffe:2020ebz, Epelbaum:2022fjc}.  Nevertheless, they provide a physically transparent access to characterize the spatial information encoded in hadronic form factors, such as radii and moments.

The Fourier transform in Eq.~\eqref{eq:Q_FT} immediately reveals the angular structure of the distribution.  We shall introduce $\widetilde{F}_Q(r) = \int\tfrac{d^3\bm\Delta}{(2\pi)^3}e^{-i\bm{\Delta}\cdot\bm r}F_Q(-\bm{\Delta}^2)$.  It is essential that $\widetilde{F}_Q(r)$ is a function of $r$ due to spherical symmetry.  We then obtain
\begin{equation}
    \Q(\bS, \br) = \bS\cdot\bm\nabla\widetilde F_Q(r) = \bS\cdot\hat{\br}{\widetilde F_Q}^\prime(r)\,,
    \label{eq:topological_dipole}
\end{equation}
where $\widetilde F_Q^\prime (r) = d \widetilde F_Q (r) / dr$.
This equation shows that the spatial distribution of the topological charge density inside the nucleon has a dipole form aligned with the nucleon spin.  This structure is robust as long as $\widetilde{F}_Q(r)$ is spherically symmetric.  Since $\Q$ is a pseudoscalar while $\widetilde{F}_Q(r)$ is a scalar, we need to construct a pseudoscalar from $\bm S$ and $\hat{\br}$, that is, their scalar product, $\bm{S}\cdot \hat{\br}$, the simplest $P$-odd combination, and therefore dictates the dominant angular structure for a spin-$1/2$ target.

The volume integration of this distribution vanishes, $\int d^3\bm r\,\Q(\br, \bS) = 0$, as is expected because the nucleon carries zero net topological charge on average.  To introduce the global measure of the distribution, we define the topological dipole moment along the spin polarization as
\begin{equation}
    d_Q = \int d^3\br\,\bS\cdot\br\,\Q(\bS, \br) \,.
    \label{eq:topological_dipole_moment_def}
\end{equation}
This quantity is given by the forward value of the topological form factor, $d_Q = - F_Q(0)$.  In the chiral limit, the axial Ward identity further predicts $d_Q = - G_A^{(0)}(0)/N_f = - \Delta\Sigma/N_f < 0$.
\vspace{0.5em}

\paragraph{Model realization:}
We quantitatively evaluate the topological dipole density in the two-flavor Skyrme model with isotriplet pseudoscalar mesons, $\pi^a$, isosinglet pseudoscalar mesons, $\eta_0$, isotriplet vector mesons, $\rho_\mu^a$, and isosinglet vector mesons, $\omega_\mu$~\cite{Jain:1987sz, Meissner:1988iv, Schechter:1999hg}.  The purpose of the model calculation is not to provide a precision determination of $F_Q(t)$, but to demonstrate the robust realization of the topological dipole structure.

Following Ref.~\cite{Schechter:1999hg}, we introduce the model Lagrangian density as
\begin{equation}
    \Lag = \Lag_0 + \Lag_1 + \Lag_V + \Lag_m + \Lag_\text{WZW} + \Lag_a\,.
\end{equation}
The nonlinear sigma term is $\Lag_0 = (f_\pi^2/4)\tr(\partial_\mu U\partial^\mu U^\dagger)$, where $U = e^{i(\eta_0 + \tau^a\pi^a)/f_\pi}$ is a $U(2)$ representation with the pion decay constant $f_\pi=\SI{93.0}{MeV}$ and the isospin Pauli matrices $\tau^a$.
The next term, $\Lag_1 = -(f_\pi^2/2)\tr[(D_\mu\xi\cdot\xi^\dagger + D_\mu\xi^\dagger\cdot\xi)^2]$ is the mixing term of pseudoscalar and vector mesons, where $\xi^2 = U$, $D_\mu = \partial_\mu - iV_\mu$, $V_\mu = g(\tau^a\rho^a_\mu + \omega_\mu)/2$.  The coupling constant is chosen as $g = 5.85$.
The kinetic term of vector mesons is $\Lag_V = -1/(2g^2)\tr(F_{\mu\nu}F^{\mu\nu})$ where $F_{\mu\nu} = \partial_\mu V_\nu - \partial_\nu V_\mu - i[V_\mu, V_\nu]$.
The pion mass term is $\Lag_m = (m_\pi^2f_\pi^2/4)\tr(U + U^\dagger - 2)$ where $m_\pi = \SI{138}{MeV}$ is the charged pion mass.
The Wess-Zumino-Witten term is $\Lag_\text{WZW} = \varepsilon^{\mu\nu\rho\sigma}\tr\bigl[\left(c_1/6 + c_2/4\right)R_\mu p_\nu p_\rho p_\sigma - i(c_2/4)F_{\mu\nu}(p_\rho R_\sigma - R_\rho p_\sigma) - (c_2 + 2c_3)R_\mu R_\nu R_\rho p_\sigma\bigr]$,
where $R_\mu = i(D_\mu\xi\cdot\xi^\dagger + D_\mu\xi^\dagger\cdot\xi)/2$ and $p_\mu = \partial_\mu\xi\cdot\xi^\dagger - \partial_\mu\xi^\dagger\cdot\xi$~\cite{Fujiwara:1984mp, Jain:1987sz}.  The coefficients $c_i$ are chosen~\footnote{For $c_i$, we follow the notation of Refs.~\cite{Jain:1987sz, Meissner:1988iv} and they are related to $\gamma_i$ used in Ref.~\cite{Schechter:1999hg} via $c_i = \gamma_i / g$  } so as to reproduce the central parameter set, $(\tilde h, \tilde g_{VV\phi}, \kappa) = (0.4, 1.9, 1.0)$, quoted in Ref.~\cite{Schechter:1999hg}.
Finally, the $U(1)_\mathrm{A}$ anomaly term is $\Lag_a = q^2/(2\topsus) + (iq/2)\ln(\det U/\det U^\dagger) - \theta q$, where $q$ is an auxiliary field representing the topological charge density in the effective theory, $\theta$ is the QCD vacuum angle, and $\topsus$ is the topological susceptibility parameter~\cite{Rosenzweig:1979ay, DiVecchia:1980yfw, Witten:1980sp, Kawarabayashi:1980dp}.

Since $q$ is nondynamical in this model, the equation of motion solves $q = \topsus(\theta + 2\eta_0/f_\pi)$, from which, at $\theta=0$, we obtain the topological charge density as
\begin{equation}
    q(\br) = \frac{2\topsus}{f_\pi} \, \eta_0(\br)\,.
    \label{eq:Q_auxiliary}
\end{equation}
The problem of computing $q(\br)$ therefore reduces to solving for the singlet pseudoscalar profile, $\eta_0(\br)$.  This relation renders the anomaly term to $\Lag_a = - (\topsus/2)(\theta + 2\eta_0/f_\pi)^2$.  This form is consistent with the well-known Witten formula to account for the mass of the singlet pseudoscalar meson, i.e., $m_{\eta_0}^2 = m_\pi^2 + 4\topsus/f_\pi^2$.  We fix $\topsus$ by fitting $m_{\eta_0}=\SI{958}{MeV}$.  

To get the profile of mesons, we first determine the static soliton in the usual hedgehog form, $U(\br) = e^{i\eta_0(\br)/f_\pi} \tilde{U}(\br)$ with $\eta_0(\br) = 0$ and $\tilde{U}(\br) = \exp[i\bm\tau\cdot\hat{\br}F(r)]$ together with the vector-meson fields $\rho^{i, a}(\br) = \varepsilon^{ika}\hat r^kG(r)/(gr)$ and $\omega^0(\br) = \omega(r)$.  The static profiles $F(r)$, $G(r)$, and $\omega(r)$ are obtained by minimizing the classical soliton energy.  For the static hedgehog, the singlet pseudoscalar field is absent because a spherically symmetric soliton has no $P$-odd structure.

We introduce the collective coordinate $A(t)\in SU(2)$ and write $A^\dagger\dot A = i\tau^aK^a$, where $\bK$ is the angular velocity.  To the first order in $\bK$, the rotating soliton is parametrized as $\tilde{U}(\br, t) = e^{i\eta_0(\br, t)/f_\pi}A(t)\tilde{U}(\br)A^\dagger(t)$ and $\rho^i(\br, t) = A(t)\rho^i(\br)A^\dagger(t)$.  The induced components of temporal $\rho$ and spatial $\omega$ are written, respectively, as $\tau^a\rho_0^a(\br, t) = (2/g)A(t)\tau^a\left[K^a\xi_1(r) + \hat{r}^aK^b\hat{r}^b\xi_2(r)\right]A^\dagger(t)$ and $\omega^i(\br, t) = [\phi(r)/r]\varepsilon^{ijk}K^j\hat{r}^k$.  The induced singlet pseudoscalar field is taken as $\eta_0(\br, t) = \eta(r)K^i\hat{r}^i$.  Substituting these Ans\"atze into the action gives the moment of inertia, $\Theta$, of the soliton.  The profiles $\xi_1(r)$, $\xi_2(r)$, $\phi(r)$, and $\eta(r)$ are then determined by minimizing $\Theta$, with the static profiles, $F(r)$, $G(r)$, and $\omega(r)$ unchanged.  The resulting Euler-Lagrange equations are solved numerically with regular boundary conditions at the origin and vanishing profiles at infinity.

Once $\eta(r)$ is obtained, the topological charge density follows from Eq.~\eqref{eq:Q_auxiliary}.  After collective quantization, the angular velocity, $\bK$, is related to the nucleon spin, $\bS$, and the topological charge density takes the following form:
\begin{equation}
    q(\bS, \br) = \frac{\topsus}{2f_\pi\Theta}\eta(r)\bS\cdot\hat{\br}\,,
    \label{eq:Q_result}
\end{equation}
which is consistent with the general dipole structure in Eq.~\eqref{eq:topological_dipole}.

\begin{figure}
    \centering
    \includegraphics[width=0.5\textwidth]{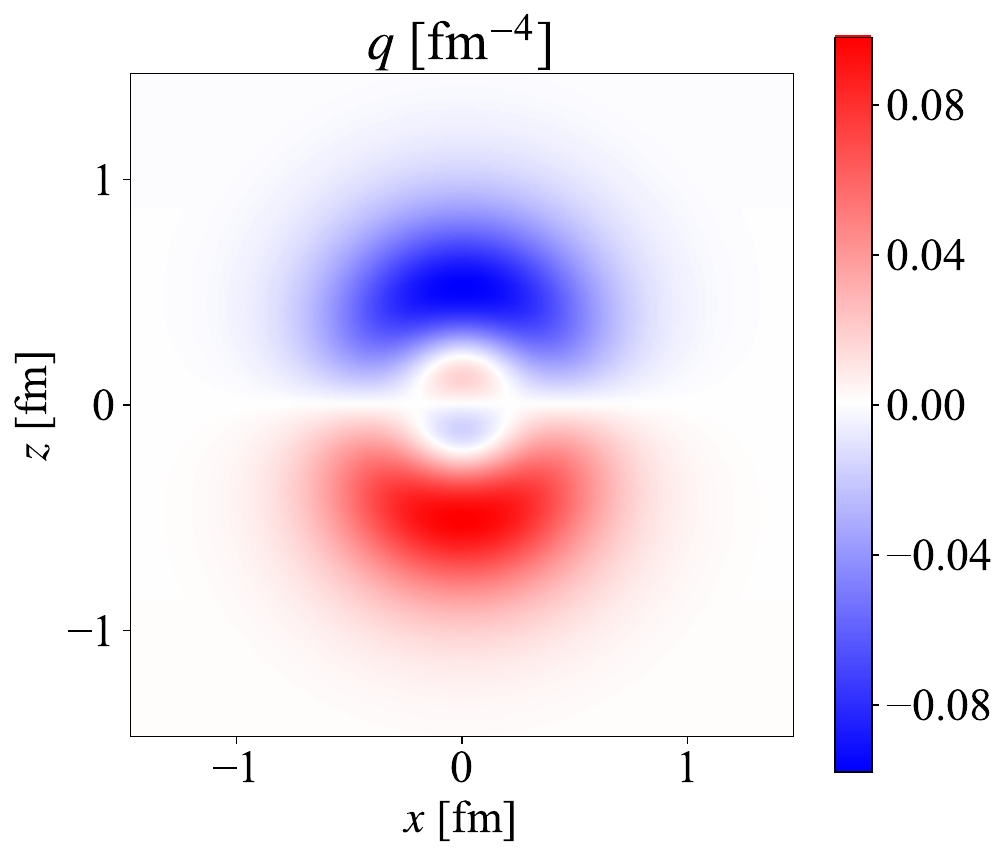}
    \caption{Spatial distribution of the topological charge density $q(\br, \bS)$ in the $x$-$z$ plane obtained from the vectorized Skyrme model.  The spin direction is chosen along the positive $z$ axis.}
    \label{fig:Q}
\end{figure}

Figure~\ref{fig:Q} shows the resulting distribution in the $x$-$z$ plane, with the spin quantization axis chosen along the positive $z$ direction.  The density exhibits a dipole pattern: it is negative in the upper half, $z>0$, and positive in the lower half, $z<0$.  This sign corresponds to a negative topological dipole moment, $d_Q = - 0.077$.  This value is smaller than the expectation from the anomalous Ward identity, namely, $d_Q\simeq -\Delta\Sigma/N_f$ with $N_f = 2$ and the phenomenological value, $\Delta\Sigma^{u+d}\sim0.4$~\cite{Cruz-Martinez:2025ahf}.  The difference is possibly attributed to the finite quark mass corrections to the anormalous Ward identity.

The sign of the dipole has a simple qualitative interpretation.  In the chiral limit, the anomalous Ward identity gives, for a static configuration, $\bm\nabla\cdot\bm{J}_5 = 2N_f\Q$.  The spatial component of the singlet axial current may be viewed as the quark-spin distribution inside the nucleon.  For a nucleon polarized along the positive $z$ direction, $\bm{J}_5$ points predominantly upward.  Since the spin density is largest near the center of the nucleon and decreases toward the outside, a volume element in the upper half receives a larger incoming axial-current flux from below than the outgoing flux above.  Hence $\bm\nabla\cdot\bm{J}_5 < 0$, which implies $\Q < 0$ for $z > 0$.  In the lower half, the situation is reversed.  The topological dipole can thus be understood from gradient patterns of the axial-current density.
\vspace{0.5em}

\paragraph{Experimental implications:}
A possible experimental probe of the topological dipole may be sought in deeply virtual pseudoscalar-meson production, $\gamma^* N \to M N^\prime$, with $M = \pi, \eta, \eta^\prime$.  Since the topological charge density in a polarized nucleon has the dipole form, $\Q(\bS, \br)\propto \bS\cdot\hat{\br}$, its sign changes under the reversal flip of the nucleon spin.  This implies that the target-spin asymmetry is a natural observable, i.e.
\begin{equation}
    A_{UT} = \frac{d\sigma(\bS_T)-d\sigma(-\bS_T)}{d\sigma(\bS_T)+d\sigma(-\bS_T)}\,,
\end{equation}
where $\bS_T$ denotes the component of the target spin transverse to the virtual-photon direction.  This expectation is also supported by the structure of pseudoscalar DVMP amplitudes.  Although $A_{UT}$ is not a direct measure of the topological charge density, it can contain interference terms involving the singlet helicity GPDs, $\widetilde H^{(0)}$ and $\widetilde E^{(0)}$~\cite{Goloskokov:2011rd}.  Their moments give the singlet axial and induced pseudoscalar form factors, $G_A^{(0)}(t)$ and $G_P^{(0)}(t)$, which constrain the topological form factor $F_Q(t)$.

In practice, however, this route is experimentally challenging and the feasibility of extracting a clean anomaly contribution is nontrivial.  The main difficulty is the separation of the helicity-GPD contribution from transverse-photon amplitudes involving transversity GPDs, $H_T$ and $E_T$, which can contribute to the same asymmetries~\cite{Goloskokov:2011rd}.  Moreover, the limited yields of $\eta$ and especially $\eta^\prime$ production make a precise extraction demanding.

Another possibility is to look for the spin-induced topological structure in the relativistic heavy-ion collision experiments.  In this case, it is not realistic to reconstruct the topological form factor of an individual nucleon with high precision.  Instead, it would be conceivable to test a spin-induced $P$-odd structure in matter.  Our observation makes a sharp contrast to the conventional arguments.  Usually, to probe local parity violation in the heavy-ion collisions, the event-averaged topological charge density itself is vanishing, while event-by-event fluctuations such as $\langle \Q(\br_1)\Q(\br_2)\rangle$, rather than $\langle \Q(\br)\rangle$, are relevant observables, where $\langle\cdot\rangle$ denotes the event average~\cite{Kharzeev:2007jp}.  By contrast, according to our proposed scenario that each nucleon carries a spin-induced topological dipole, an external electromagnetic field can provide a preferred axial direction and induce a nonzero $\langle \Q(\br)\rangle$ as well as the spin alignment.

In noncentral collisions, the spectator protons generate a strong magnetic field, $\bB$, approximately perpendicular to the reaction plane~\cite{McLerran:2013hla}.  We assume that participant nucleons and stopped baryons acquire a nonzero event-averaged spin polarization correlated with $\bB$.  Alternatively, the polarization could be induced by the mechanical rotation with the angular velocity vector $\bm{\omega}$.  Thus, we use a notation, $\bBo$ to mean either $\bB$ or $\bm{\omega}$ or their mixture.  Since each baryonic constituent carries a local topological dipole, the medium can be viewed as an ensemble of dipoles with $\bBo$-biased orientation, which is analogous to a magnet as a result of microscopic magnetic dipoles.  After taking the event average, we can no longer resolve the individual dipole but observe the collective manifestation of a coarse-grained topological polarization correlated with $\bBo$.  For further discussions, let us adopt the simplest idealization, assuming that the coarse-grained polarization in matter has a similar dipole structure, which depends only on the radial distance from the center of the fireball.  This assumption gives the global dipole form as
\begin{equation}
    \langle \Q(\br)\rangle_{\bBo} = P_Q(r)\hat{\bBo}\cdot\hat{\br}\,,
    \label{eq:Q_evant_averaged}
\end{equation}
where $P_Q(r)$ encodes the radial profile of the coarse-grained topological polarization.  To discuss the observability of Eq.~\eqref{eq:Q_evant_averaged}, it is convenient to introduce a source field, $\Theta_\text{eff}(\br)$, conjugate to $\Q$.  It would be a natural assumption that this background field has the same dipole structure as the topological charge density, and we parametrize the angular dependence of this effective background as
\begin{equation}
    \Theta_\text{eff}(\br) \propto \hat{\bBo}\cdot\hat{\br}\,,
    \label{eq:Theta_effective}
\end{equation}
which encodes the event-averaged $P$-odd structure of the medium.  The effective background $\Theta_\text{eff}$ and $\theta_0$ couple through an anomaly-induced interaction, $\Lag_\text{int}\sim g_{\eta}\eta_0\Theta_\text{eff}$.  This interaction causes a source for the field configuration of $\eta_0$, as a solution of the equation of motion,
$(\Box + m_{\eta_0}^2)\eta_0 \sim g_\eta \Theta_\text{eff}$.

The corresponding production amplitude is proportional to the on-shell Fourier component of the source, representing a dipole harmonic in momentum space.  The amplitude for $\eta$ and $\eta^\prime$ production is therefore proportional to
\begin{equation}
    \mathcal M_{\text{odd}}[\eta_0(\bp)] \:\:\propto\:\: g_\eta \tilde{\Theta}_\text{eff} (\bp) \:\:\propto\:\: i \hat{\bBo}\cdot \hat{\bp}_T\,.
\end{equation}
In the setup of the heavy-ion collision, the magnetic field is approximately perpendicular to the reaction plane, and the transverse momentum of the emitted meson is measured in the plane perpendicular to the beam axis.  Therefore, for particles emitted in the transverse plane, the angular pattern of the production amplitude is given by $\sin(\phi - \Psi_\text{SP})$, where $\phi$ is the azimuthal angle of the emitted meson and $\Psi_\text{SP}$ denotes the spectator-plane angle.

It is important to note that $\mathcal M_{\text{odd}}$ has a definite phase associated with real source fields, $\Theta_\text{eff}(\br)$, and the $i$ factor arises from the Fourier transform of the dipole structure.  This phase is crucial for the interference with the ordinary $P$-even production amplitude, $\mathcal M_\text{even}$, which is dominated by a resonance such as $\eta^\prime(958)$.

Interference with the ordinary $P$-even production amplitude, $\mathcal{M}_\text{even}$, then gives the azimuthal yield distribution of $\eta$ and $\eta^\prime$ mesons,
\begin{equation}
\begin{split}
    \frac{dN_{\eta, \eta^\prime}}{d\phi} &\sim |\mathcal{M}_\text{even} + \mathcal{M}_\text{odd}|^2  \\
    &\simeq \frac{dN_0}{d\phi}\left[1 + 2a_1^{\eta,\eta'}\sin(\phi - \Psi_\text{SP}) + \cdots\right]\,.
\end{split}
\end{equation}
The eclipsed dots denote higher harmonics, and the first-harmonic coefficient $a_1^{\eta,\eta'}$ is read from the expansion of the interference term, $\Re(\mathcal{M}_\text{even}^*\mathcal{M}_\text{odd})/|\mathcal{M}_\text{even}|^2$.  The sign of $a_1^{\eta,\eta'}$ is determined by the relative phase between the $P$-even and $P$-odd amplitudes.  In particular, if the $P$-even amplitude is dominated by a resonance, such as $\eta^\prime(958)$, the phase is expected to be close to $\pi/2$, which gives a fixed phase in the interference term.
In this way, the topological dipole would manifest itself as a directed-flow-like azimuthal asymmetry of $\eta$ and $\eta^\prime$ mesons correlated with the spectator plane.

The above argument is robust regardless of concrete shapes of $\Theta_\text{eff}$, and should be contrasted with the conventional CME scenario~\cite{Kharzeev:2007jp}.  There, the sign of the topological charge generated by sphaleron transitions fluctuates event by event.  The CME current is therefore parallel or antiparallel to the magnetic field, depending on this sign~\cite{Fukushima:2008xe}.  As a result, the observable quantities must always be even under the $P$-transformation, which in a harmonic language naturally points to second-harmonic correlations, rather than to a $v_1$ signal.
\vspace{0.5em}

\paragraph{Summary and outlook:}
We have proposed a nontrivial distribution of the topological charge density in a spin polarized nucleon.  The pseudoscalar nature of this phenomenon leads to a dipole form, i.e., $\Q(\bS, \br)\propto \bS\cdot\hat{\br}$.  

For a demonstration purpose, we showed a chiral soliton calculation with the $U(1)_A$ anomaly, which explicitly realizes this topological dipole pattern.  Such a structure would provide a direct evidence for the local $P$-odd topological structure along the spin axis of a nucleon.  We have discussed two possible experimental implications: exclusive $\eta$, $\eta^\prime$ production and magnetic-field-correlated azimuthal asymmetries.  Alternatively, from the theoretical point of view, it would be intriguing to diagnose the topological distribution using the lattice QCD simulations.  In principle, the topological form factor, $F_Q(t)$, should be a measurable observable in the lattice QCD and it can quantify the strength of the topological dipole.  Further investigation of the topological dipole in both theory and experiment would provide a new window to explore the interplay between QCD topology and spin physics.
\vspace{0.5em}

\acknowledgments
The authors thank
Yui~Hayashi,
Yan~Lyu,
Christian~Weiss, and
Ismail Zahed
for useful discussions.
This work was supported by Japan Society for the Promotion of Science
(JSPS) KAKENHI Grant Nos.\ 
22H05118, 25K24464, 26K00698 (K.F.) and
FoPM, WINGS Program, The University of Tokyo (T.U.).

\bibliographystyle{apsrev4-2}

\bibliography{topological_dipole}
\end{document}